\documentclass[aps,showpacs,prb,twocolumn,supperscriptaddress]{revtex4-1}
\usepackage{graphicx} 
\usepackage{dcolumn} 
\usepackage{bm} 
\usepackage{graphicx,color,epsfig,rotate}
\usepackage{amsmath,amssymb}
\usepackage{pifont}

\makeatletter
\newcommand{\Rmnum}[1]{\expandafter\@slowromancap\romannumeral #1@}
\makeatother

\begin{document}
\title{Monte Carlo simulations of vector pseudospins for strains: Microstructures,\\  and martensitic conversion times}
\author{N.~Shankaraiah}
\affiliation{School of Physical Sciences, Jawaharlal Nehru University, New Delhi 110067, India.}

\date{\today}

\begin{abstract}
We present systematic temperature-quench Monte Carlo simulations on discrete-strain pseudospin model Hamiltonians to study microstructural evolutions in 2D ferroelastic transitions with two-component vector order parameters ($N_{OP}=2$). The zero value pseudospin is the single high-temperature phase while the low-temperature phase has $N_v$ variants. Thus the number of nonzero values of pseudospin are triangle-to-centered rectangle ($N_v=3$), square-to-oblique ($N_v=4$) and triangle-to-oblique ($N_v=6$). The model Hamiltonians contain a transition-specific Landau energy term, a domain wall cost or Ginzburg term,  and power-law anisotropic interaction potential, induced from a  strain compatibility condition. On quenching below a transition temperature, we find behaviour similar to the previously studied square-to-rectangle transition ($N_{OP} =1, N_v = 2$), showing that the rich behaviour found, is generic. Thus we find for two-component order parameters, that the same Hamiltonian can describe both athermal and isothermal martensite regimes for different material parameters. The athermal/isothermal/austenite parameter regimes and temperature-time-transformation diagrams are understood, as previously, through parametrization of effective-droplet energies.   
In the athermal regime, we find rapid conversions below a spinodal like temperature and austenite-martensite conversion delays above it, as in the experiment. The delays show early incubation behaviour, and at the transition to austenite the delay times have Vogel-Fulcher divergences and are insensitive to Hamiltonian energy scales, suggesting that entropy barriers are dominant.   \end{abstract}

\maketitle

\section{ Introduction}

Steels and shape memory alloys are martensitic materials that undergo diffusionless, first-order phase transformation from high-temperature parent 'austenite' unit-cell to low-temperature product 'martensite' unit-cells (or variants) on cooling or under external stress \cite{R1}. A subset of physical strains are the order parameter (OP). As martensitic materials have many applications \cite{R1, R2}, much work has been done to understand domain-wall microstructures and their underlying kinetics. According to traditional classification \cite{R3}, martensites are classified as {\it athermal}, with rapid milli-second austenite-martensite conversions on cooling below a martensite start temperature and no conversions above it; and {\it isothermal}, which can have slow or delayed conversions in minutes or hours. But, experiments on athermal martensitic materials have found delayed conversions above the martensite start temperature, where only austenite should exist \cite{R4}. Computer simulations of martensitic models could give insights into the classification of martensites and the unexpected delayed-conversions in athermal martensites.

Continuous variable nonlinear free energies are minimized in displacement, phase field, and strain using relaxational dynamics, Monte Carlo (MC) and Molecular dynamics simulations \cite{R5, R6, R7, R8, R9} and the obtained microstructures are consistent with experiment \cite{R10}, but that can need extensive computer time. More economic discrete-strain  clock-like model Hamiltonians \cite{R11} are systematically derived from continuous strain free energies for different ferroelastic transitions in 2- and 3-spatial dimensions (2D \& 3D). Power-law anisotropic interaction potentials, which arise from the no-defect St.Venant compatibility condition \cite{R11, R12}, and orient strain domain walls, have their counterparts induced in pseudo spin Hamiltonians. The microstructures generated from these strain-pseudospin models using local mean-field approximation \cite{R13} are in good agreement with the continuous variable models \cite{R5, R6, R7} and experiments \cite{R10}. 

Systematic temperature-quench MC simulations were performed on the simplest  scalar-OP, 3-state pseudospin Hamiltonian for square-to-rectangle (SR) transition \cite{R14}  and showed both rapid conversions below a spinodal-like temperature and incubation-delays above it, as in experiments \cite{R4} on athermal martensitic materials. The conversion-time delays found to have Vogel-Fulcher divergences, which are insensitive to Hamiltonian energy scales and log-normal distributions, suggesting the dominant role of entropy barriers.  An athermal/isothermal martensites regime diagram is predicted in material-parameters; crossover temperatures and domain-wall phases in Temperature-Time-Transformation (TTT) diagrams are understood through parametrization of textures by surrogate  droplet energies; and role of power-law potentials are shown to be important for textures and incubations \cite{R14}. 
The central question is: Are such conversion-delays in the athermal martensite regime, specific to the scalar-OP transition, or are they generic, appearing in vector-OP transitions ?

In this paper, we show that the athermal martensite regime conversion-delays in the scalar-OP ($N_{OP}=1$) SR transition are generic in three vector-OP ($N_{OP}=2$) ferroelastic transitions: triangle-centered rectangle $(N_{v}=3)$; square-oblique $(N_{v}=4)$; triangle-oblique $(N_{v}=6)$. Under systematic MC temperature quenches, we find isothermal parameter regime with slow or delayed conversions and athermal parameter regime that has rapid conversions below a temperature and incubation-delays above it, as in experiment \cite{R4} and scalar-OP SR transition \cite{R14}. The athermal regime conversion-time delays have Vogel-Fulcher divergences, which are insensitive to Hamiltonian energy scales and log-normal distributions. The athermal/isothermal/austenite regime diagrams are obtained in material parameters. The crossover temperatures and domain-wall phases in the TTT diagram are understood through the parametrization of textures. Microstructures obtained in these transitions are in good agreement with continuous-variable simulations \cite{R5, R6, R7} and experiment \cite{R10}. We finally show the importance of power-law interaction potentials in the incubation behaviour,  and microstructures. 

The paper is organised as follows. In Section 2, we outline derivations of the vector-OP strain-pseudospin Hamiltonians. In Section 3, we present the athermal/ isothermal martensite regimes and crossover in material parameters. In Section 4, we focus on the athermal martensite regime and present conversion-delay kinetics, parametrization of domain-wall phases in TTT diagram by effective droplet energies, and conversion incubation textures. In Section 5, we present kinetics in the absence of the power-law anisotropic interactions that shows delays without incubation, and Section 6 is a summary.

\section{Strain-pseudospin hamiltonians}

In this Section, we state for completeness, the vector-OP strain-pseudospin model Hamiltonians \cite{R11}, that were systematically derived from scaled continuous-strain free-energies \cite{R12} for triangle-to-centered rectangle (TCR), square-to-oblique (SO) and triangle-to-oblique (TO) ferroelastic structural transitions.

In 2D, structural transitions have  $d (d + 1)/2 =3$ or three distinct physical strains, the compressional ($e_1$), deviatoric ($e_2$) and shear ($e_3$) strains. Of these, ($e_2, e_3$) are OP ($N_{OP}=2$) and the $e_1$ is non-OP ($n=1$) strains. The scaled free energy has a Landau term ${\bar F}_{L}$; a Ginzburg term, quadratic in the OP gradients ${\bar F}_G $; and a seemingly innocuous term, quadratic in the non-OP strains ${\bar F}_{non} $, that turns out to generate crucial power-law anisotropic interactions between the OP strains \cite{R11}. Thus 
 $$F = E_0 [ {\bar F}_L + {\bar F}_G + {\bar F}_{non} ]~~ (2.1)$$
Here $E_0$ is an elastic energy per unit cell.
The transition specific Landau term $\bar F_{L}$ has ($N_{v}+1$) degenerate energy minima at the first-order transition as shown in Figure 1. The high-temperature austenite minima is allowed at all temperatures as its existence has to be determined dynamically, and $N_{v}$ minima are the low-temperature martensite variants. The pseudospin derivation results of Ref 11 are restated here, for completeness. 

\begin{figure}[h]
\begin{center}
\includegraphics[height=2.7cm, width=8.2cm]{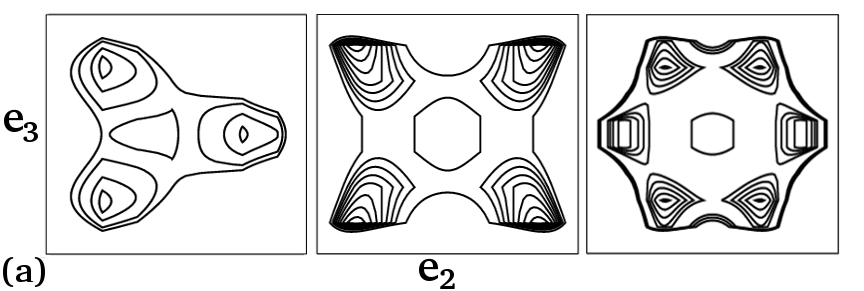}
\includegraphics[height=2.8cm, width=8.4cm]{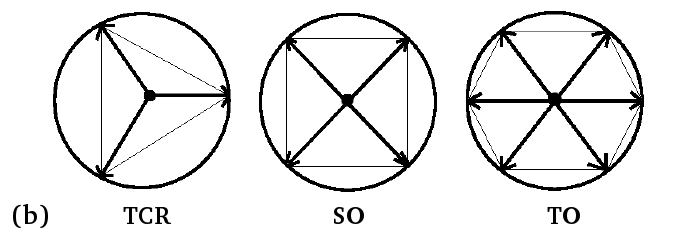}
\caption{{\it Landau free energy minima and strain-pseudospin clock vectors:} (a) Contours of Landau free energy $\bar F_{L}$ showing single austenite minima at center and $N_v$ martensite minima at corners, in a plot of $e_3$ versus $e_2$ and (b) corresponding clock model minima at pseudospin values for TCR ({\it left}), SO ({\it middle}) and TO ({\it right}) transitions.}
\end{center}
\label{Fig.1}  
\end{figure}

The scaled Landau free energy ${\bar F}_L $ for TCR transition \cite{R11}
$${\bar F}_L  (\vec e) =\sum_{\vec r}  (\tau - 1) {\vec e}^{2} + [\vec e^{2}-2(e_2^3-3e_2 e_3^2)+(\vec e^2)^2], ~~(2.2)$$
has an austenite minima at $(e_2 , e_3) = (0 , 0) $, and  $N_v = 3$ martensite minima at which  $(e_2 , e_3)= (\cos \phi , \sin\phi)$ for $ \phi=0 , \frac{2\pi}{3}, \frac{4\pi}{3}$.
Here ${\vec e}^2 \equiv e_2 ^2 + e_3 ^2$, and $\tau=(T-T{c})/(T_{0}-T_{c})$ is the scaled temperature; $T_{0}$ is the first-order Landau transition temperature and $T_{c}$ is metastable austenite spinodal temperature.

The scaled Landau free energy ${\bar F}_L $ for SO transition \cite{R11}
$${\bar F}_L  (\vec e) =\sum_{\vec r}  \tau {\vec e}^{2} - (4-C^{'}_{4}/2){\vec e}^{4}+4{\vec e}^{6}-C^{'}_{4} e_{2}^{2} e_{3}^{2}, ~~(2.3) $$
also has an austenite minima at $(e_2 , e_3) = (0 , 0) $, and  $N_v = 4$ martensite minima at which  $(e_2 , e_3)= (\cos \phi , \sin\phi)$ for $ \phi=\frac{\pi}{4} , \frac{3\pi}{4}, \frac{5\pi}{4} $ and $ \frac{7\pi}{4}$
with material dependent elastic constant $C^{'}_{4}$.

The scaled Landau free energy for TO transition \cite{R11} is
$${\bar F}_L  (\vec e) =\sum_{\vec r}  (\tau - 1) {\vec e}^{2} + {\vec e}^{2}({\vec e}^{2}-1)^{2}+C_{6}({\vec e}^{3}-(e_2^3-3e_2 e_3^2)^{2}), ~(2.4)$$
where $C_{6}$ is a material dependent parameter. The Landau polynomial has an austenite minima at $(e_2 , e_3) = (0 , 0) $, and  $N_v = 6$ martensite minima at which  $(e_2 , e_3)= (\cos \phi , \sin\phi)$ for $ \phi=0, \frac{\pi}{6} , \frac{2\pi}{6}, \frac{3\pi}{6}, \frac{4\pi}{6} $ and $ \frac{5\pi}{6}$.

The domain-wall cost Ginzburg term $\bar F_{G}$ is,
$$ {\bar F}_G (\vec \nabla \vec e) = \xi^2 \sum_{\vec r} (\vec \nabla {\vec e})^2 ~~~(2.5) $$

The non-OP term is harmonic \cite{R11}, with stiffness $A_1$,
$${\bar F}_{non} (e_1)  = \sum_{\vec r} \frac{A_1}{2} e_1 ^2 = \sum_{\vec k} \frac{A_1}{2} |e_1|^2~(2.6)$$
and is minimized subject to St.Venant compatibility constraint for physical strains \cite{R12},

 $$ \Delta ^2 e_1 - (\Delta_x ^2 - \Delta_y ^2) e_2 - 2\Delta_x \Delta_y  e_3 = 0;~~(2.7a)$$ 
with gradient terms as difference operators ${\vec \nabla} \rightarrow {\vec \Delta}$ for sites $\vec r$ on a computational grid. 
In Fourier space $k_\mu  \rightarrow K_\mu (\vec k) \equiv 2 \sin (k_\mu /2)$ and so
$$ O_1 e_1 + O_2 e_2 + O_3 e_3 =0 ~~ (2.7b)$$
where the coefficients for square lattice are  $O_1=-\frac{1}{\sqrt{2}}{\vec K } ^2$, $O_2=\frac{1}{\sqrt{2}}({ K_x ^2 - K_y ^2 }) $, and $O_3={2K_x  K_y }$; for triangle lattice, $O_1=-{\vec K } ^2 $, $O_2=({ K_x ^2 - K_y ^2 })$, and $O_3={2K_x  K_y }$. Here, $\vec K^2=(K^2_x+K^2_y)$.
Minimization of non-OP strain generates power-law anisotropic interactions between OP strains, by inserting a direct solution $e_1 = - (O_2 e_2 + O_3 e_3)/ O_1$ for  $\vec k \neq 0$ into (2.6),
$$ F_{compat} (e_2,e_3 ) = \frac{A_1}{2} \sum_{\ell, \ell^{'}=2, 3; \vec k} e_{\ell}(\vec k) ~U_{\ell \ell^{'}}(\vec k) ~e^{\ast}_{\ell^{'}}(\vec k) ~(2.8)$$
where $ U_{22} (\vec k) =  \nu (O_2/O_1) ^2$ , $U_{23} (\vec k) =  \nu (O_2 O_3/O_1) ^2$ , $U_{33} (\vec k) =  \nu  (O_3/O_1) ^2$ and $\nu=(1-\delta_{\vec k,0})$. Figure 2 shows the power-law potentials $U_{\ell\ell^{'}}$ as relief plots in Fourier space and contours in coordinate space.

\begin{figure}[h]
\begin{center}
\includegraphics[height=5.5cm, width=8.5cm]{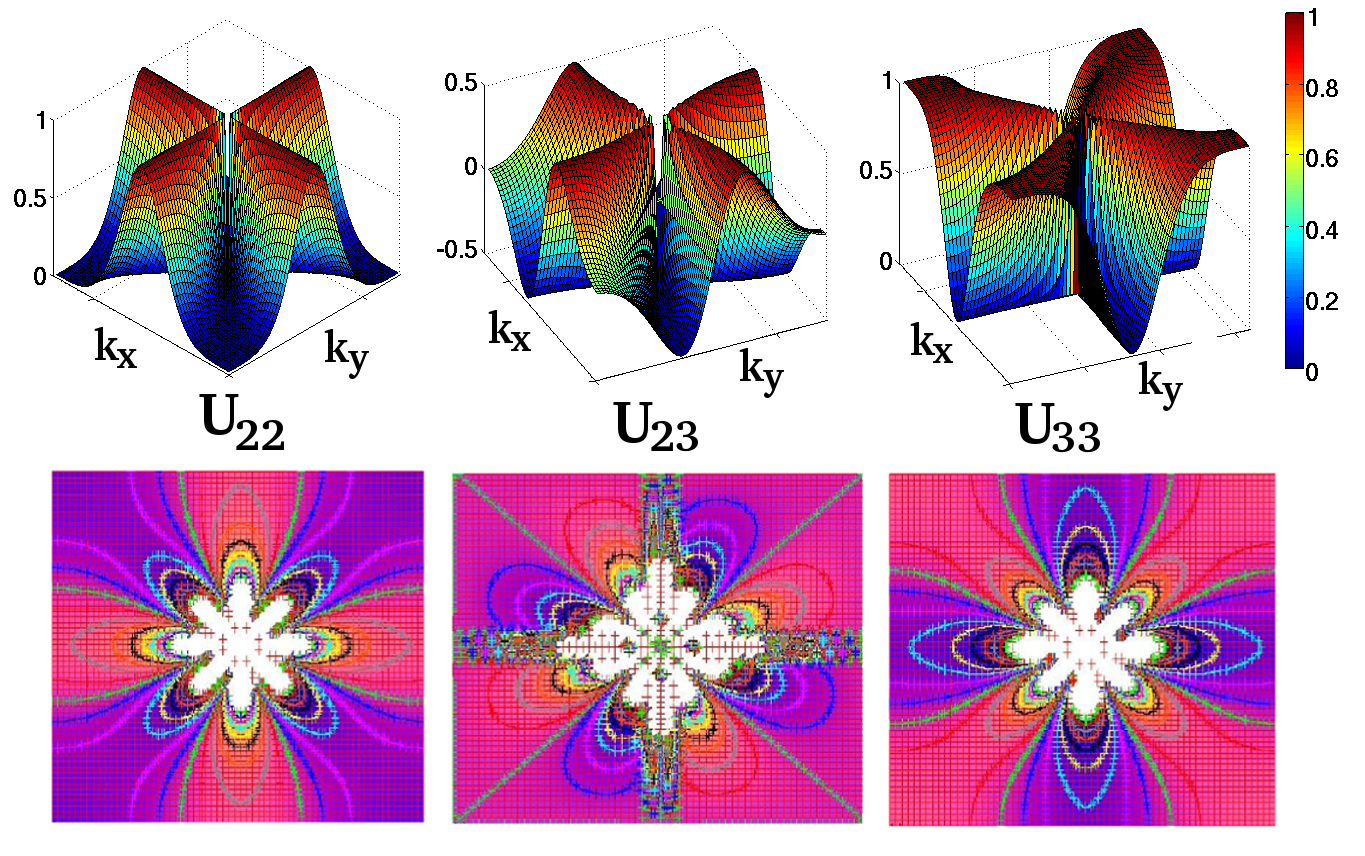}
\caption{{\it Power-law anisotropic potentials:} Relief plots of kernels $U_{22}(\vec k)$, $U_{23}(\vec k)$, and $U_{33}(\vec k)$ in Fourier space ({\it top row}) and corresponding contours in coordinate space ({\it bottom row}).}
\end{center}
\label{Fig.2}
\end{figure}

The continuous-strain OP $\vec e= (e_2, e_3)$ is discretized \cite{R11} by choosing its values only at the $N_{v}+1$ Landau minima,
$$\vec e (\vec r) = |e| (\cos\phi , \sin\phi) \rightarrow  {\bar \varepsilon }(\tau) \vec S(\vec r). ~~  (2.9) $$ 

The Landau term \cite{R11} becomes,
$$ H_L  (\vec S)=  {\bar \varepsilon}^2 \sum_{\vec r} g_L (\tau)  {\vec S}^2(\vec r)= {\bar \varepsilon}^2 \sum_{\vec k} g_L (\tau) |\vec S(\vec k)|^2 ~~(2.10)$$
where $g_L = \tau - 1  + ( \bar \varepsilon - 1)^2, {\bar \varepsilon}^2 (\tau) =\frac{3}{4} \{ 1 + \sqrt{ 1 - 8 \tau /9} \}$ for TCR, and $g_L = \tau - 1  + ( {\bar \varepsilon}^2 - 1)^2, {\bar \varepsilon}^2 (\tau) =\frac{2}{3} \{ 1 + \sqrt{ 1 - 3 \tau /4} \}$ for SO and TO transitions.

The square gradient Ginzburg term becomes,
$$ H_G (\vec \nabla \vec S) = \xi^2  \sum_{\vec r} {\bar \varepsilon}^2 (\vec \nabla {\vec S})^2  = \xi^2 \sum_{\vec k} K^2 {\bar \varepsilon}^2 |{\vec S}(\vec k)|^2~~~(2.11) $$

The discrete-strain pseudospin clock-zero model Hamiltonian is derived \cite{R11} by substituting (2.9) into the total free energy (2.1),
$$\beta H( \vec S) \equiv \beta { F} (\vec e \rightarrow {\bar \varepsilon } \vec S )~ (2.12)$$

The Hamiltonian in coordinate space is 
$$\beta H = \frac{D_0}{2}[  \sum_{{\vec r} } \{ g_L(\tau) {\vec S}^{2} (\vec r) +    \xi^2 {(\vec \Delta \vec S)}^2   \}$$
$$+   \sum_{{\vec r}, {\vec r'}} \sum_{\ell, \ell^{'}=2,3} \frac{A_1}{2} U_{\ell \ell^{'}} ({\vec r} - {\vec r'} ) S_{\ell} (\vec r) S_{\ell{'}}  (\vec r') ],~(2.13a)$$
where $D_0= 2E_0 {\bar \varepsilon}^2$.
It is diagonal in Fourier space,

$$\beta H = \frac{1}{2} \sum_{\vec k} \sum_{\ell,\ell^{'}=2,3}  Q_{0, \ell \ell^{'}} (\vec k) S_{\ell} (\vec k) {S^{\ast}_{\ell^{'}}} (\vec k)  ;~~(2.13b)$$
$$Q_{0, \ell \ell^{'}} (\vec k) \equiv D_0 [\{ g_L(\tau) +  \xi^2 { \vec K}^2  \}\delta_{\ell,\ell^{'}}  + \frac{A_1}{2} U_{\ell \ell^{'}} ({\vec k})],~(2.13c)$$
and is a clock-zero model Hamiltonian with single austenite $\vec S=(S_2,S_3)=(0,0)$ and $N_v$ martensite variants: 
$$\vec S= (1 , 0) , (-\frac{1}{2} , \pm \frac{\sqrt 3}{2});   (\pm \frac{1}{2} , \pm \frac{1}{2}); (\pm 1,0), (\pm \frac{1}{2}, \pm \frac{\sqrt 3}{2})~(2.14) $$
for TCR  ($N_{v}+1=4$), SO ($N_{v}+1=5$), and TO ($N_{v}+1=7$) transitions respectively.

\begin{figure}[h]
\begin{center}
 \includegraphics[height=10.0cm, width=8.5cm]{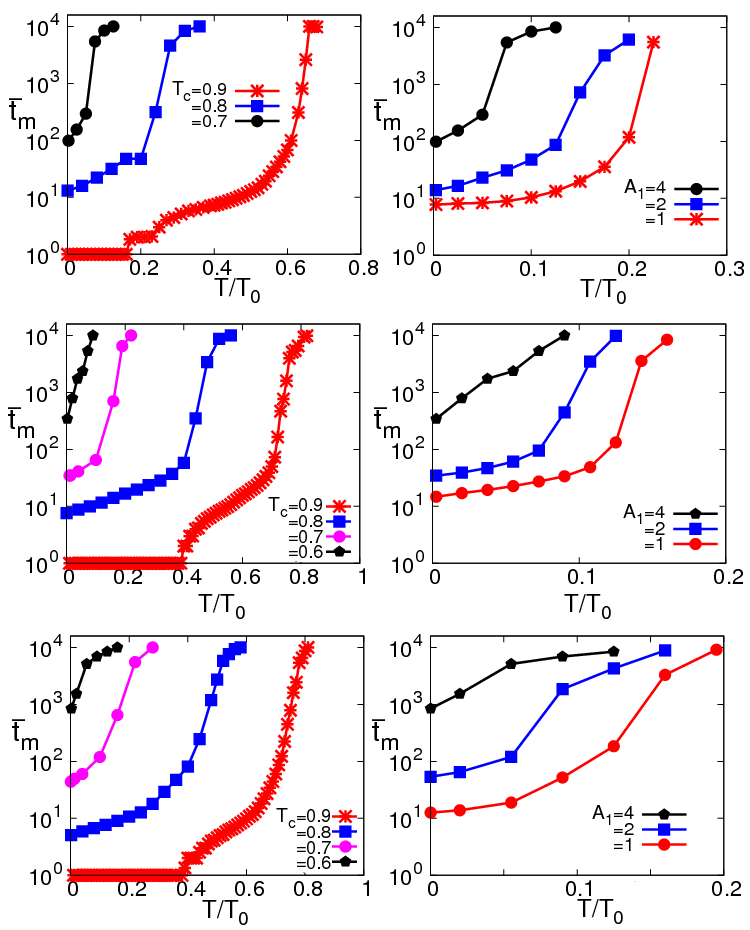}
\caption{{\it Crossover behaviour of  athermal/isothermal martensite conversions:} {\it Left:} Conversion times $\bar t_m$ vs $T/T_0$ showing crossover from athermal (fast) to intermediate (slow) for fixed $E_0=3, A_1=4$ and $T_c=0.9,0.8,0.7$. {\it Right:} Conversions showing crossover from intermediate (slow) to athermal (fast) for fixed $E_0=3, T_c=0.7$ and $A_1=4,2,1$. {\it Top row:} TCR, {\it middle row:} SO, and {\it bottow row:} TO transitions.}
\end{center}
\label{Fig.3}
\end{figure}
MC temperature-quench simulations are carried out systematically \cite{R14} on a square lattice in 2D. At $t=0$, we consider $2\%$ of sites randomly with $N_v$ strain-pseudospin martensite values in austenite.  The seeds are quenched below the Landau transition $\tau<<1$ and held for $t \le t_h$ MC sweeps (MCS). Metropolis algorithm is used for acceptance of energy changes, that are calculated through Fast Fourier transforms. In each MC sweep, we visit all $N=L \times L$ sites randomly, but only once. Simulation parameters are $L=64, T_0 = 1$; $T_c /T_0 = 0.6, 0.7, 0.8, 0.9$,  $\xi = 1; A_1 = 1, 4, 10; 2 A_1 / A_3 = 1$; $E_0 =3, 4, 5, 6; t_h \leq 10,000$  sweeps, and conversion times are averaged over $N_{runs} = 100$ runs.

\begin{figure}[h]
\begin{center}
 \includegraphics[height=15.0cm, width=6.5cm]{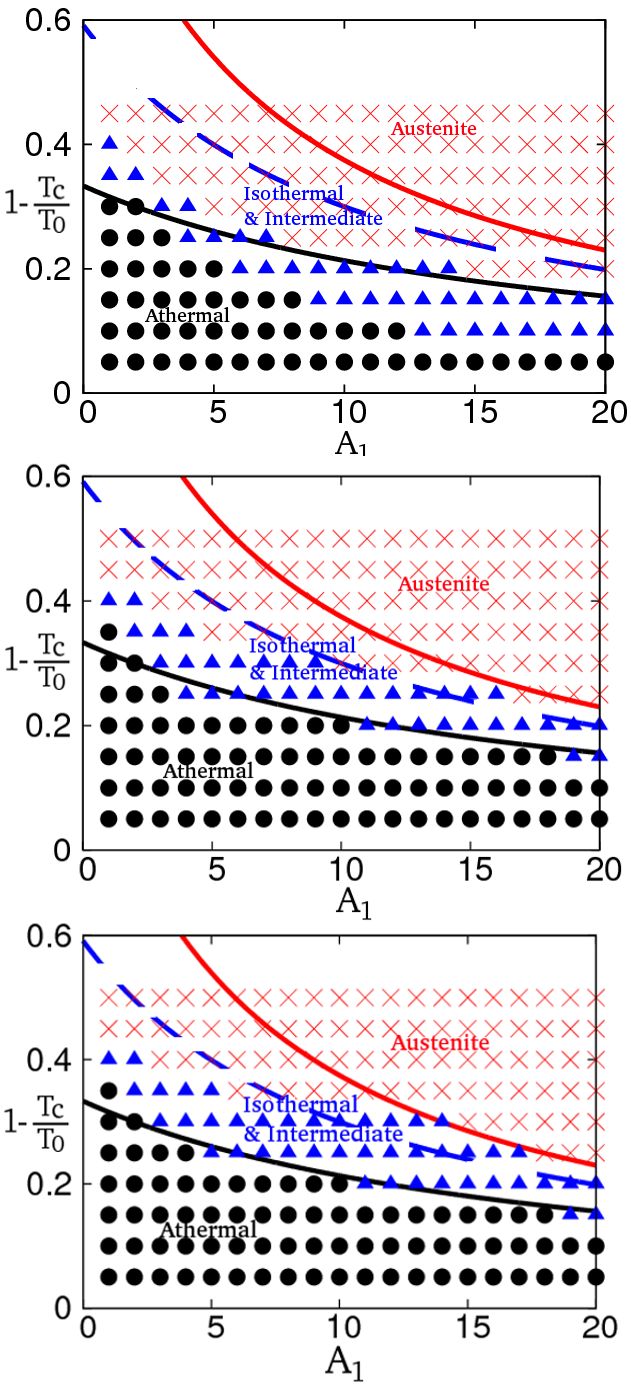}
\caption{{\it Athermal/isothermal martensite and austenite phase regimes:} Data of athermal, intermediate, and austenite behaviour in a plot of $1- \frac{T_c} {T_0}$ versus $A_1$, with $E_0 =3$ for TCR ({\it top}), SO ({\it middle}) and TO ({\it bottom}) transitions. Estimates of the theoretical boundaries are shown as solid lines. See text.}
\end{center}
\label{Fig.4}
\end{figure}

\section{Athermal and isothermal parameter regimes}

On quenching $2\%$ of martensite seeds to a temperature $\tau(T) <  \tau(T_0)$, we define \cite{R14} martensite conversion fraction $n_m(t)$, which is equal to 0 in the pure austenite and 1 in the pure/twinned martensite,
$$ n_m(t)=\frac{1}{N} \sum_{\vec r} {\vec S}^2({\vec r}), ~~(3.1) $$
and specify conversion time $t = t_m$ when $n_m (t_m) =  0.5$.  

From Figure 3, we can see isothermal slow conversions and athermal fast conversions with incubation-delay tails for different material parameters $A_1, T_c/T_0$ in TCR, SO, and TO transitions. Figure 3 also shows crossover from athermal to isothermal by fixing $A_1$ and changing $T_c/T_0$, and vice versa. Hence, we find the martensite classification is a matter of material parameters: the same model Hamiltonian can show both athermal or isothermal behaviour, dependent on parameters. This is just as in the SR case\cite{R14}. 
 
\begin{figure*}
\begin{center}
 \includegraphics[height=4.0cm, width=17.0cm]{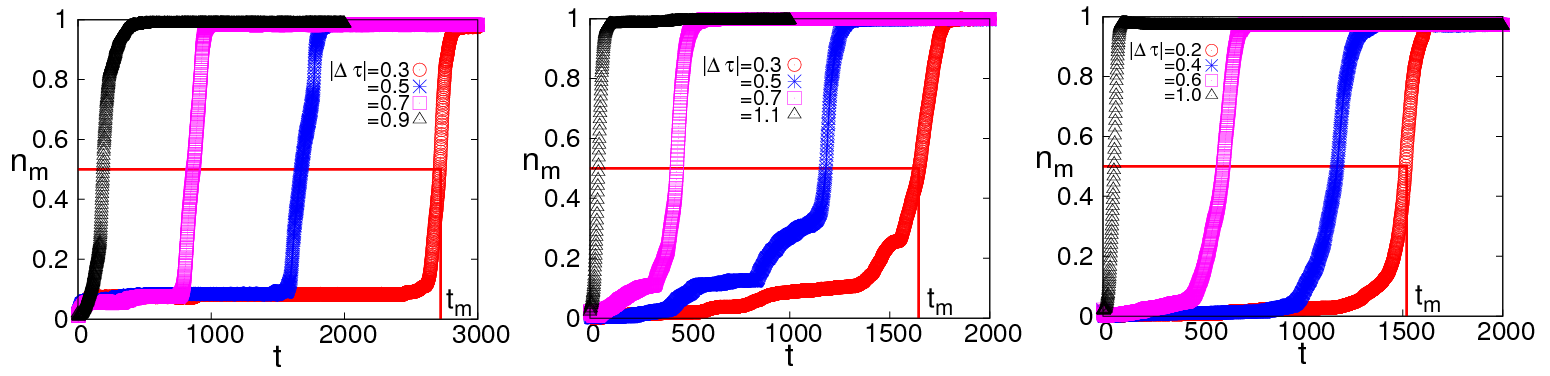}
\caption{{\it  Conversion incubation times:} Martensite conversion fraction $n_m (t)$ versus time $t$, for stiffness $A_1 =4$ and various $\Delta \tau \equiv \tau - \tau_4 < 0$, showing the $50 \%$ conversion definition of $t_m$ for TCR ({\it left}), SO ({\it middle}) and TO ({\it right}) transitions. As athermal regime parameters have been chosen, flat incubation is seen for early times. }
\end{center}
\label{Fig.5}
\end{figure*}

The athermal/isothermal/austenite regime diagrams are obtained in material parameters $T_c/T_0, A_1$ and shown in Figure 4, that clearly depicts athermal martensites are more common than isothermal \cite{R2}.  The simulations data matches well with the estimates \cite{R14} of theoretical boundaries. Here, the criterion for athermal is $\bar t_m=10$ MCS; isothermal/intermediate is $\bar t_m=1000$ MCS; and austenite, if there are no conversions even for holding time $t=t_h$. Again this is just as in the SR case \cite{R14}.

We will focus on the athermal regime. Figure 5 shows single-seed runs of $n_m (t)$ vs $t$ after  quenches to various $\Delta \tau\equiv \tau  - \tau_4 < 0$, below the transition temperature $\tau_4 \equiv \tau (T_4)$. At low temperatures, $n_m (t)$ rises rapidly to unity, but as transition is approached, shows {\it  incubation} behaviour. In the case of TCR transition, $n_m(t)$ rises to a smaller value, that incubates for longer times before it rises sharply to unity. In SO transition, $n_m(t)$ shows incubation followed by jerky steps before it rises to unity. In TO transition, $n_m(t)$ has longer incubations before it sharply rises to unity. The transition is 'fuzzy' and is operationally defined as where all 100 runs give austenite. Hence, we define \cite{R14} mean conversion time $\bar t_m=1/<r_m>$, where mean conversion rate $<r_m>=<1/t_m>$ is obtained by an arithmetic average over $N_{runs}=100$ seeds.

We henceforth focus on the athermal martensite parameter regime to study the conversion-delays kinetics.

\section{Textural energies  parametrized by surrogate droplets}

The transition is known to depend both on temperature and the size of martensitic seeds, as in the Pati-Cohen model \cite{R3}. In early work, Pati and Cohen \cite{R3} have measured and modeled the conversion times in Ni-Mn alloys and found that the isothermal slow conversions can change to athermal fast conversions, for fixed martensite fraction, but with larger (and hence fewer) initial martensite seeds.
This can be understood through the parametrization of textural droplet energies as in SR transition \cite{R14}. 
 At $t=0$, the seeds of $N_{v}$ variants are randomly sprinkled throughout the lattice. We find that the interaction tend to cancel leaving only self-interaction part $A_1 [U]/2$ at each seed. So we have, 
$$\beta H[\vec S(0)] \simeq \frac{D_0}{2} \sum_{\vec r}[ g_L  {\vec S}({\vec r}, 0)^2 + \xi^2 \{{\vec \Delta} {\vec S}({\vec r}, 0)\}^2 $$
$$+ \frac{A_1 [U]}{2} {\vec S}(\vec r,0)^2]. ~~(4.1)$$
Here, $[U] \simeq 0.5$, is the Brillouin-zone average of $U_{\ell \ell^{'}}(\vec k)$ of (2.8) in TCR, SO, and TO transitions. For an initial  martensite fraction $n_{m} (0) =0.02$, we have $N_v$ variants square seeds of sides $R(0)$. The initial  pseudospin seed energy is parametrized as  $\beta H(\vec S(0)) =  C_0 [ \alpha_L g_L R(0)^2 + \alpha_G \xi^2 4 R(0) +  \alpha_C (A_1 [U]/2) R(0)^2]$ with $C_0 \equiv (n_m(0) N D_0/2)$.   For different sides $R(0) = 1, 2, 3$, we fit the coefficients $\alpha_{L, G, C}$ term-by-term, finding again $\alpha_L = \alpha_G = \alpha_C  =1$, independent of seed size. Then  the initial energy has a droplet-like form $\beta H(R(0)) = C_0 2 \xi^2 R_c [1 - (1- R(0)/R_c)^2]$. Here we define a length  $R_c (\tau)$ that is positive below a divergence temperature $\tau = \tau_L (A_1)$, 
$$R_c (\tau) \equiv  \frac{-2 \xi^2}{g_L (\tau) + A_1 [U]/2}. ~~(4.2)$$ 
As in the SR case, we define a scaled temperature variable $\eta(\tau)$ from the parametrization  
$$\eta (\tau) = - R(0)/ R_c (\tau) = \frac{g_L (\tau) + A_1 [U]/2}{2 \xi^2}, ~~(4.3)$$
that will be used later, for $R(0) =1$.

At $t=0$, the initial seeds have a geometric meaning, and hence the pseudospin Hamiltonian $H(\vec S(0))$ matches the droplet Hamiltonian $H(R(0))$, but for general $t$,  these two terms no longer match term-by term. However, as in SR case, we define $R(t)$ through  $H[\vec S(t)]/ H[\vec S(0)] =H[R(t)]/ H[R(0)]$.  The energy (ratio) for interacting {\it vector} pseudospins is parametrized, by the energy (ratio) of a surrogate system  of independent droplets.  The initially geometric  $R(0)$ evolves to  an interacting-texture energetic parameter $R(t)$, that  can even go negative as the pseudospin energy goes negative.
Thus
$$\rho(t) \equiv \beta H(\vec S(t))/ \beta H(\vec S(0)) $$
$$=  [1 - (\frac{R(t)}{R_c} - 1)^2]/[1 - (\frac{R(0)}{R_c} - 1)^2]. ~~(4.4)$$

The $R(t)$ evolution is then once again
$$R(t)/R_c = 1 +  \alpha \sqrt{ 1 - \rho (t) /\rho_c}, ~~(4.5)$$
where  $\rho_c \equiv [ 1 - (\{R(0)/R_c \}- 1)^2 ]^{-1}$, and we take $\alpha = sign (\{R(0)/ R_c\} - 1).$

\begin{figure}[h]
\begin{center}
 \includegraphics[height=2.5cm, width=8.5cm]{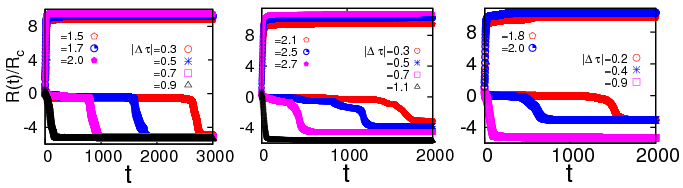}
\caption{ {\it Trajectories :} Scaled energy parameter $R(t)/R_{c}$ versus time t, showing flows are determined by initial $R(0)/R_{c} (\tau)$ values. Note  flat incubations of lower curves, of initial  $1> R(0)/R_{c} (\tau) >0.5$, corresponding to $\tau_{2} < \tau < \tau_{4}$ for TCR ({\it left}), SO ({\it middle}) and TO ({\it right}) transitions.} 
\end{center}
\label{Fig.6}
\end{figure}

Figure 6 shows the evolutions of effective droplet energy in a plot of  $R(t)/R_c (\tau)$ versus time. There are both rapid rises to final positive values and flat-incubations as already seen in the martensite conversion fraction $n_m$, which goes negative at later times. The flat-incubations are due to the inefficient searches for the rare channels to lower energies. The initial $R(0)/R_c$ values determine the $R(t)$ flows.

As a consistency test of  parametrization, Figure 7 shows  $\rho (t) / \rho_c$ versus $R(t)/R_c$  indeed matches a parabola,  for all $t$, and all $A_1$, and many starting values  $R(0)/ R_c (\tau)$ in TCR, SO, and TO transitions. Flow directions  of $R(t)$ are indicated by arrows starting at $R(0)/R_c$ for Regions 1,2,3,4, with asymptotic $R(t)$ giving negative final martensitic energies, or zero (going to austenite).

\begin{figure}[h]
\begin{center}
 \includegraphics[height=6.0cm, width=8.0cm]{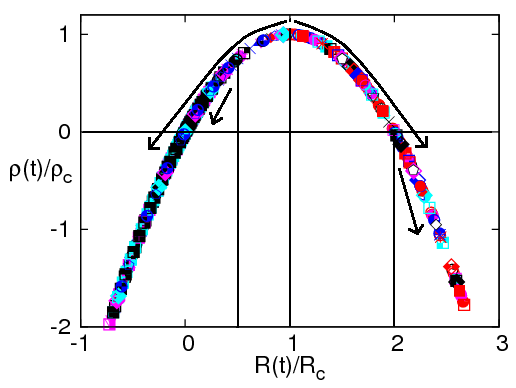}
\caption{{\it Parametrization and crossover  temperatures :} Scaled pseudospin energy  $\rho (t)/\rho_c$ versus $R(t)/R_c (\tau)$, showing flows fall on a parabola as a test of parametrization. For $R(0) =1$ seeds, characteristic  initial values  $R(0)/R_c$  are ${R_{c1}}^{-1} =2, {R_{c2}}^{-1} = 1,  {R_{c4}}^{-1} \simeq 0.5$ as marked. These correspond to temperatures $\tau_1, \tau_2, \tau_4$.  For initial $R(0)/ R_{c} \lesssim 0.5$, flows are to $R=0$ austenite.} 
\end{center}
\label{Fig.7}
\end{figure}

i) {\it Region 1:} For initial $R(0)/R_c(\tau) > 2$,  there are explosive conversions to martensite, this determines a temperature  $\tau = \tau_1$ or $1/ R_c (\tau_1) = 2$ or in scaled varibale $\eta(\tau)=-2$ with initial unit seeds $R(0) = 1$.  

ii) {\it Region 2:} For  initial droplets  $2> R(0)/R_c(\tau) >1$,  the flows are again fast, this  determines a temperature $\tau = \tau_2$ where $1/R_c(\tau_2)=1$ or $\eta(\tau)=-1$.  

iii) {\it Region 3:} For $0.5 \gtrsim R(0)/R_c(\tau) $ or $\eta(\tau) \gtrsim -0.5$, the initial droplets are flowing  only to $R=0$ austenite. But, for larger $A_1$, the droplets can still grow through searches up to $R(0)/R_c(\tau) \simeq 0$ or $\eta(\tau) \simeq 0$, that is well below the Landau transition temperature $T_0$.

iv) {\it Region 4:} For  $0.5 \lesssim R(0)/R_c(\tau)  \lesssim 1$, the initial droplets immediately convert to a single variant droplet, that incubates for long times around $R(t) \simeq 0$ with zero energy $H \simeq 0$ (degenerate with austenite). This entropically critical droplet searches for conversion pathways, and grows through jerky steps and autocatalytic twinning. The incubations occur for unit seeds up to a temperature $\tau = \tau_4$ or $1/R_c (\tau_4)  \simeq 0.5$ or in scaled variable $\eta(\tau)=-0.5$. 

These critical values of the scaled variable $\eta (T, A_1)$ are used in the scaled plot of Fig. 8.
\begin{figure}[h]
\begin{center}
\includegraphics[height=6.0cm, width=8.3cm]{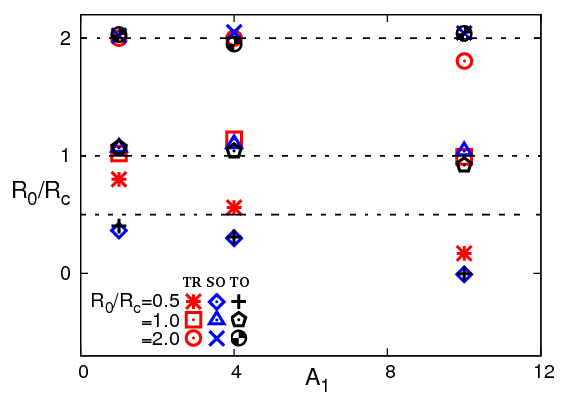}
\caption{{\it  Textural crossover temperatures :} Phase diagram $R_{0}/R_{c}$ vs $A_1$. Theoretical boundaries (shown as dotted lines) in  $\tau_1, \tau_2,  \tau_4$  are defined by $R_c (\tau_{1,2,4}) = R_{c1, c2, c4}$ of Fig 7 and symbols are data from simulations. See text.}  
\end{center}
\label{Fig.8}
\end{figure}

\section{Athermal regime conversions}

In this Section, we find the conversion times and their distributions with a data collapse in terms of a scaled temperature-stiffness variable; and textural evolutions.

\subsection{Conversion time and their distributions}

\begin{figure*}
\begin{center}
\includegraphics[height=5.0cm, width=18.0cm]{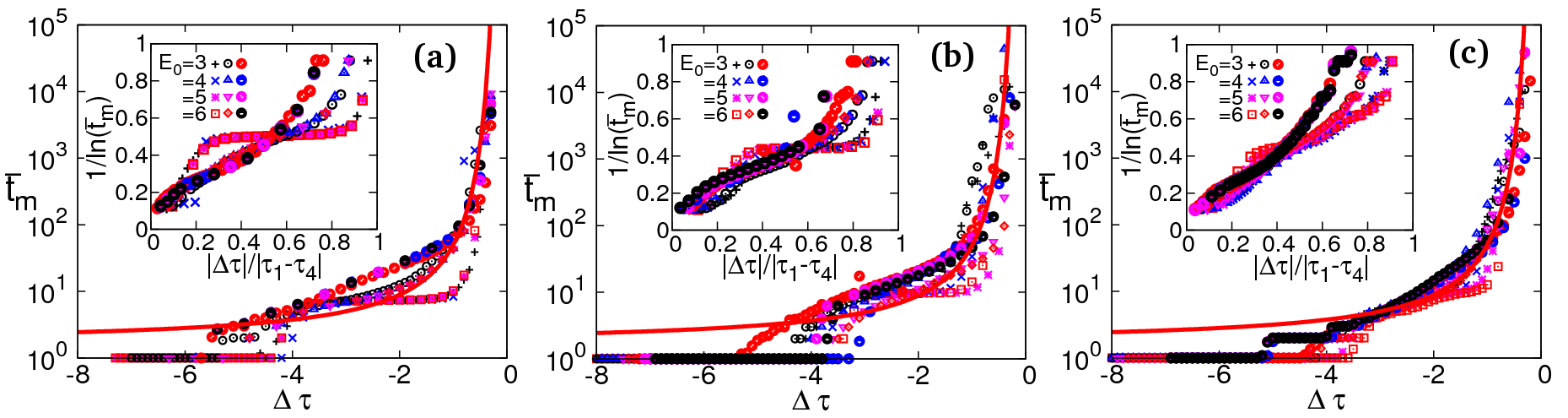}
\caption{{\it Singular divergence of conversion times:} Plot of $\ln {\bar t}_m$ vs $|\Delta  \tau| \equiv |\tau - \tau_4|$, for ({\it a}) TCR , ({\it b}) SO and ({\it c}) TO transitions with $T_c /T_0 =0.9,  E_0 =3,4,5,6$ and $A_1=1, 4, 10$. Solid line is  $\ln {\bar t}_m = \ln t_0  + b_0 |\tau_1 - \tau_4|/|\Delta \tau|$, with $t_0 = 1.6$ and $b_0 = 1.7$. }
\end{center}
\label{Fig.9}
\end{figure*}
The conversion times for different material parameters $A_1, E_0$ and for TCR, SO, and TO transitions fall on a single hyperbola in $\log {\bar t}_m$ versus $\Delta \tau \equiv  \tau - \tau_4$,  for a range of temperatures $\tau_4 < \tau < \tau_1$ as shown in main Fig 9. The same data is plotted in inset as $1/\ln({\bar t}_{m})$ versus $|\Delta \tau|/ |\tau_1-\tau_4|$, that shows linearity on $|\Delta \tau| $ goes to zero. The hyperbola and the linearity are showing the Vogel-Fulcher behaviour \cite{R15}. Specifically,  ${\bar t}_{m} = t_0 \exp [ b_0 |\tau_1 - \tau_4|/|\{ \tau  - \tau_4\}|]$, with $t_0 = 1.6, b_0 = 1.7$, for these data.
The insensitivity of conversion times ${\bar t}_{m}$ to energy scales $E_0$ implies that the Vogel-Fulcher behaviour at $\tau_4$ comes from divergence of entropy (rather than energy)  barriers, ${\bar t}_m \sim e^{|\Delta S_{entr}|}$ in finding the rare channels \cite{R16}. The entropy barriers then vanish at $\tau_1$, with a drop in conversion times.

The main Figure 9 shows the temperature dependence of conversion times \cite{R17}  for TCR, SO, and TO transitions. As in the scalar-OP SR transition, there are explosive conversions below a temperature $\tau \simeq \tau_1$ (that is different for different transitions); and there are conversion delays for $\tau \gtrsim \tau_1$, that rise at $\tau \simeq \tau_2$, to diverge at a temperature $\tau \simeq \tau_4$.
The success conversion fraction $\phi_m$ versus $\Delta \tau$  for various $A_1 = 1, 4, 10$ and $E_0=3,4,5,6$ with fixed $T_c=0.9$ for TCR, SO, and TO transitions is shown in Supplemental Material. The fraction $\phi_m$ is unity for $\tau \lesssim \tau_2$ and decreases for $\tau > \tau_2$, to become zero at $\tau \simeq \tau_4$. The insensitivity to different energy scales $E_0$ is again a signature of entropy barriers \cite{R16}.

\begin{figure}[h]
\begin{center}
\includegraphics[height=2.5cm, width=8.7cm]{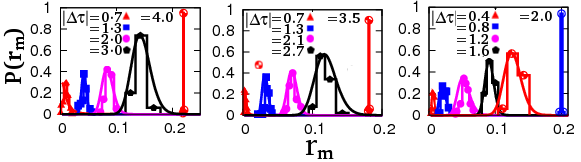} 
\caption{{\it  Log-normal distribution  of conversion-rates:}  Plot of $P(r_m)$ versus $r_m$ for different $\Delta \tau$. Data are from Fig.9, with $A_1=4,E_0 =3$ for TCR ({\it left}), SO ({\it middle}) and TO ({\it right}) transitions. Solid lines are the log-normal distributions.} 
\end{center}
\label{Fig.10}
\end{figure} 

Similar to the SR transition, we have calculated the arithmetic mean rate $< r_m> \equiv <1/t_m>$ that determines ${\bar t}_m= 1/<r_m>$, with $1/t_h < r_m < 1$ in TCR, SO, and TO transitions. The variance in the rates  is ${\sigma_r}^2 = <( r_m -< r_m>)^2>$.  The probability densities $P(r_m)$ versus $r_m$ for various $\Delta \tau$ are shown  in the Figure 10, as histograms for different temperatures. For each histogram of $N_{hist}$ data points, the Scott optimized bin size \cite{R18}  is used, of $dr_m = 3.5 \sigma_r /[N_{hist}]^{1/3}$. The histograms again narrow sharply, below $\tau_2$, as in the delta-function-like  peak on the right. Solid lines are Log-normal curves for calculated $<r_{m}>$ and $\sigma_{{r}_{m}}^2$ from the data. The Log-normal distribution is a signature of rare events \cite{R19}.   

\begin{figure}[h]
\includegraphics[height=6.5cm, width=8.5cm]{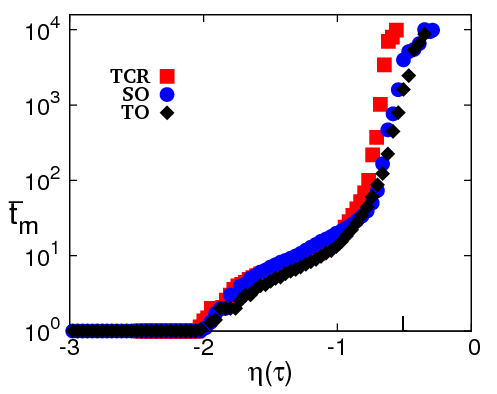}
\caption{{\it Temperature-Time-Transformation phase diagram :} Conversion times ${\bar t}_m$  versus  scaled variable $\eta(\tau)$ for TCR, SO, and TO transitions showing fast and delayed conversions for given parameters $A_1 = 4, E_0 =3,T_c /T_0 = 0.9$.}
\label{Fig.11}
\end{figure}

\begin{figure*}
\begin{center}
\includegraphics[height=4.9cm, width=17.0cm]{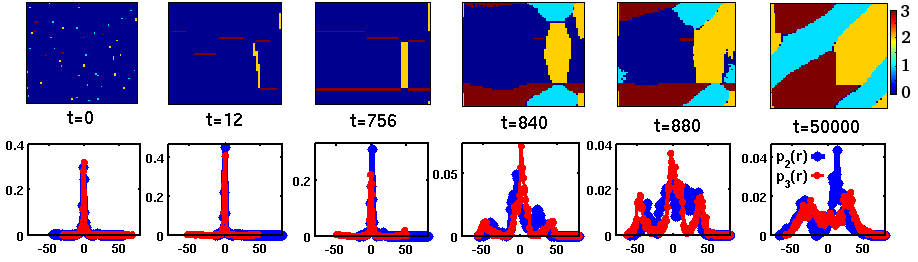}
\caption{{\it  Evolution for the Triangle-centered rectangle (TCR) transition: } {\it First row:} Snapshots of OP strain $\vec S=(S_2,S_3)$ for different times $t$ on quenching to $\tau=-2.7$. See movie of this evolution. The color bar is in terms of variant label $V$. See text. {\it Second row}: Evolving stress distributions of $p_{2}(\vec r), p_{3}(\vec r)$. Parameters: $A_1=4$, $E_0=3$.}
\end{center}
 \label{fig.12}
\end{figure*}

\begin{figure*}
\begin{center}
\includegraphics[height=4.9cm, width=16.7cm]{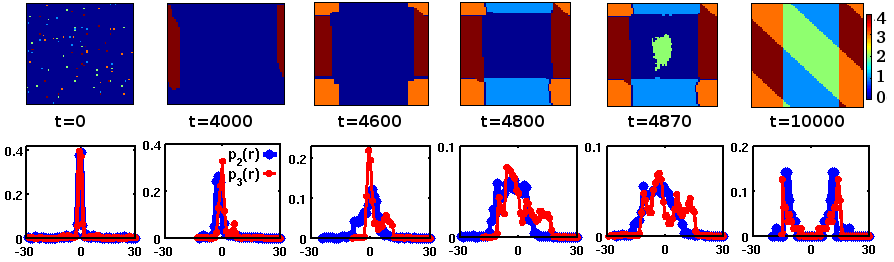}
\caption{{\it  Evolution for the Square-oblique  (SO) transition:} {\it First row:} Snapshots versus time $t$, of OP strain $\vec S=(S_2,S_3)$ on quenching to $\tau=-1.1$. The colour bar is in terms of variant label $V$. See movie.  {\it Second row:}  Evolving stress distributions $p_{2}(\vec r), p_{3}(\vec r)$. Parameters: $A_1=4$, $E_0=3$.}
\end{center}
\label{fig.13}
\end{figure*}

\begin{figure*}
\begin{center}
\includegraphics[height=4.9cm, width=16.7cm]{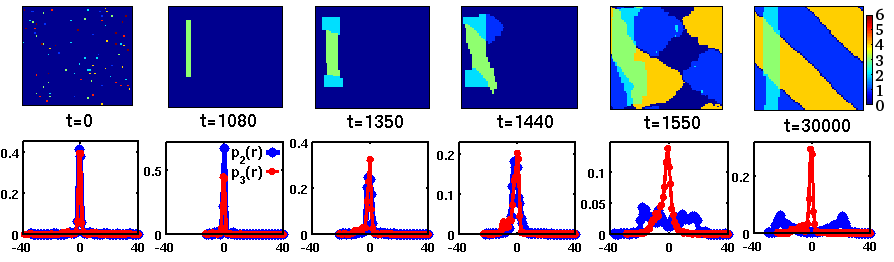}
\caption{{\it  Evolution for the  Triangle-oblique (TO)  transition:} {\it First row:} Snapshots of the OP strain $\vec S=(S_2,S_3)$ evolution in terms of $V$ on quenching to $\tau=-1.1$. See movie. {\it Second row:} Stress distributions $p_{2}(\vec r), p_{3}(\vec r)$. Parameters: $A_1=4$, $E_0=3$.}
\end{center}
 \label{fig.14}
\end{figure*}
Figure 11 shows TTT phase diagram in conversion times $\bar t_m$ versus scaled variable $\eta(\tau)=-1/R_c(\tau)$ for fixed $A_1=4$, $E_0=3$ for TCR, SO, and TO transitions.  
The characteristic temperatures $\tau_{1,2,4}$ are defined in terms of scaled variable as $\tau=\tau_1$ or $\eta(\tau)=-2$ where $\bar t_m \simeq 1$ MCS; $\tau=\tau_2$ or $\eta(\tau)=-1$ where $\bar t_m \simeq 10$ MCS; and $\tau=\tau_4$ or $\eta(\tau)=-0.5$, where conversion times diverges.

\section{Evolution of  textures}

In the athermal parameter regime, after quenching into $\tau_4 > \tau > \tau_2$, we monitor evolution of OP strain $\vec S$ textures, local internal stresses (See Appendix.) and stress distributions to understand the conversion-incubations at microstructure level. The color bar in Figs 12, 13, 14, 15 represents the OP strain $\vec S$ in terms of variant label $V$. In all the three transitions, $V=0$ represents austenite $\vec S=0$ and $V = 1,2...N_v$ corresponds to $N_v$ martensite variants with pseudospin vector values given in (2.14), and pictured in Fig 1.  

As shown in Figs. 12, 13, 14 (first row), after quenching the austenite with $2\%$ martensite seeds into $\tau_4 > \tau > \tau_2$, the seeds quickly form domain-wall 'vapour' of droplets of single variant(s), reminding Ostwald ripening. The droplets searches for the rare pathway channels to expand in the easy directions of the anisotropic potential. The expanded droplet then generates the other variant by autocatalytic twinning as in Bales and Gooding \cite{R7, R11} to form 'liquid' of domain-walls, which orient to form domain-wall 'crystal' at a later time. The jerkiness during conversion incubation is reflected in wavenumber $k_m$ (not shown) as steps with finite values \cite{R14, R20} and also in (excess) thermodynamic quantities \cite{R14, R20}, internal energy $\Delta U$ and entropy $-T \Delta S$ (not shown).
 
In the second row of Figs. 12, 13, 14, the local stress distributions $p_2(\vec r), p_3(\vec r)$ are shown. At $t=0$, the stress distributions are sharply peaked around zero with large values, which generate wings on both sides of the peak during autocatalytic twinning. In the final oriented state, only the wings remain, that correspond to the trapped stress values along the domain-walls (except in TO case, where $p_3(\vec r)$ is sharply peaked around zero). 

\begin{figure}[h]
\begin{center}
\includegraphics[height=1.8cm, width=8.65cm]{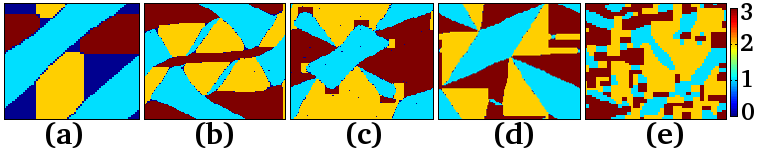}
\includegraphics[height=1.8cm, width=8.7cm]{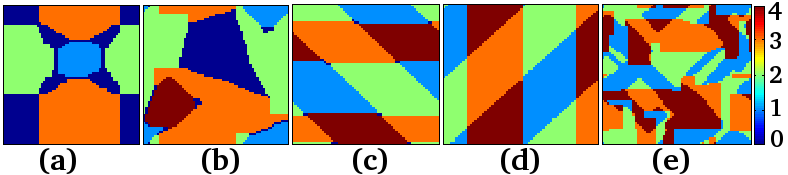}
\includegraphics[height=1.8cm, width=8.65cm]{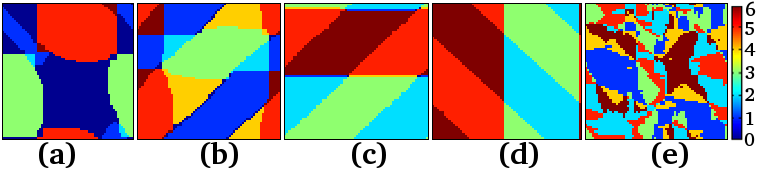}
\caption{{\it Final 'equilibrium' microstructures:}  The long time textures are shown, for  transitions  TCR, SO, and TO  are in the first, second and third rows, respectively. (a) precursor vibrating phase (bulk-austenite) for $\tau \simeq \tau_4$; (b) and (c) fan-like and Z-shaped twins (austenite appears as point or line densities) for $\tau_2 \gtrsim \tau \gtrsim \tau_4$; (d) star-like or martensite twins (no bound austenite) for $\tau_1 \gtrsim \tau \gtrsim \tau_2$; (e) glass for $\tau<< \tau_1$. The austenite is represented as zero in color bar. Parameters are typically $A_1=4, T_c=0.9, \xi^2=1$ and $E_0=3$.}
\end{center}
\label{Fig.15}
\end{figure}
The final 'equilibrium' microstructures in the TTT phase diagram for TCR, SO, and TO transitions are shown in Figure 15 and are in good agreement with continuous variable simulations \cite{R5, R6, R7} and experiments \cite{R10}.  

With random initial seeds, there is a vibrating martensite phase, that has bulk austenite in TCR, SO, and TO transitions as in Fig.15 (a) , that could be equivalent of the chequerboard SR case tweed pattern \cite{R14} (and becomes less probable closer to $\tau \simeq \tau_4$.).

With $2\%$ of martensite seeds, and for $ \tau > \tau_4$ or $\eta(\tau) \gtrsim -0.5$, there is only uniform austenite. For $\tau_2 \lesssim \tau \lesssim \tau_4$ or $-1 \lesssim \eta \lesssim -0.5$, there are again austenite droplets but now appear as lines, in domain wall crystal (DWC) in SO, TO cases, and Z-like states \cite{R10} in TCR case as in Fig.15 (b). For $\tau_1 \lesssim \tau < \tau_2$ or $-2 \lesssim \eta(\tau) < -1$, austenite droplets can appear as points at corners, in DWC (that have topological charges) in SO, TO cases; and also fan-like oriented states \cite{R10} in TCR case as shown in Fig.15 (c).  
At low temperatures for $\tau << \tau_1$ or $\eta(\tau) << -2$,  the DWC or oriented twins can have vortex-like (or topological defects or charges) behaviour at multi-variant junctions in SO \cite{R10}, TO cases, and partially oriented star-like states \cite{R10} in TCR case, can compete with a frozen domain wall liquid or 'glass' as shown in Fig.15 (d), (e), and has no bound austenite. Hence, $\tau_1$ is austenite (local) spinodal temperature.
The microstructure (d) as shown in Fig.15 (first row) for TCR transition is not fully relaxed even after $t_h=10^6$ MCS and could possibly take longer and longer times, to orient fully to a nested star as seen in continuous variable simulations and experiments \cite{R5,R6,R7,R10}.

\section{Conversions without the Compatibility interaction}

We now turn-off the compatibility term ($A_1=0$) in the Hamiltonians for TCR, SO and TO transitions to understand the role of the power-law anisotropic potentials in the domain-wall conversion kinetics.

\begin{figure}[h]
\begin{center}
\includegraphics[height=3.2cm, width=8.5cm]{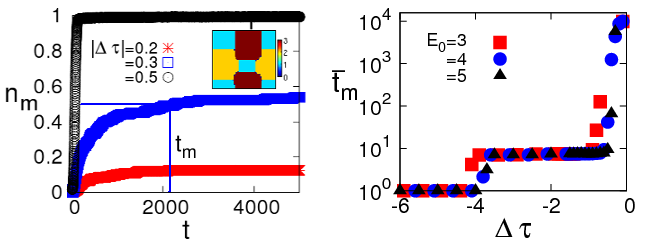}
\includegraphics[height=3.3cm, width=8.5cm]{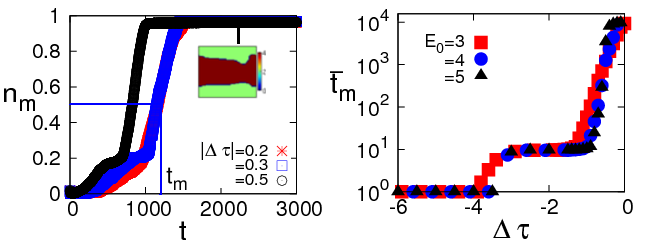}
\includegraphics[height=3.3cm, width=8.5cm]{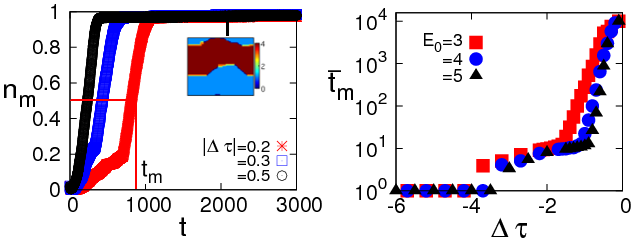}
\caption{{\it Microstructures and conversion times with $A_1=0$:} The left column shows martensite fraction $n_m(t)$ versus $t$ for $|\Delta \tau|=0.2, 0.3, 0.5$ and $E_0=3$, with the conversion time $t_m$ is marked for TCR ({\it top row}), SO ({\it middle row}), and TO ({\it bottom row}). The right column shows the  conversion times  $\bar t_m$ versus $\Delta \tau$ for $E_0=3, 4, 5$. Compare to Fig 5 and Fig. 9, respectively.} 
\label{Fig.16} 
\end{center}
\end{figure}

Figure 16 showing the martensite fraction $n_m (t)$ versus time $t$ for different temperatures $|\Delta \tau|=|\tau-\tau_4|$, where $\tau_4$ is austenite transition temperature; and coversion times $\bar t_m$ versus temperature $\Delta \tau$ for different energy scales $E_0$ for TCR, SO, and TO transitions with elastic stiffness constant $A_1=0$. Here, $n_m$ {\it has no flat regions or incubation} as was seen for $A_1=4$ in Fig.5. The final microstructure in $A_1=0$ case is a slab-like martensite unlike oriented microstructures in $A_1=4$ as in Fig.15.  

Conversion times show a rise at $\tau_1$, that almost remain constant till $\tau_2$ and then increase linearly to $\tau_4$, above which there are no conversions found. There are no Vogel-Fulcher rises in conversion times as in $A_1=4$ (Fig.9), but there is a small $E_0$ dependence, which could be now from energy barriers rather than entropy barriers.

With  the parametrization scale of (4.2) now given by $R_c (A_1 =0)=2 \xi^2/|g_L|$, the estimated transition temperatures are $\tau_1=-5.5,-4.0,-4.0$; $\tau_2=-2.0,-1.4,-1.4$; $\tau_4=-0.4,-0.2,-0.2$ are in good agreement with the simulation values of  $\tau_1=-5.5,-4.0,-4.0; \tau_2=-2.0,-1.4,-1.4; \tau_4 \simeq -0.8,-0.1,-0.1$ for TCR, SO, and TO transitions respectively. 

Therefore, microstructures and conversion times in TCR, SO, and TO transitions with $A_1=0$ are clearly different from the non-zero $A_1$ case. Thus, the power-law anisotropic potentials in ferroelastic transitions are important in understanding orientations and kinetics of domain-walls.

\section{Summary and further work}

Systematic temperature-quench MC simulations without extrinsic disorder are carried on the strain pseudospin clock-zero model Hamiltonians, with vector-order parameter ($N_{OP}=2$) and $ N_{v}+1 ( =4, 5, 7$) strain-pseudospins, that correspond to triangle-to-centred rectangle, square-to-oblique, and triangle-to-oblique transitions to get insights into conversion-incubation kinetics. The results are similar to the SR case \cite{R14}, that are just seen to be generic. The simulation results are as follows: 

(1) The microstructures of discrete-strain pseudospins in the Temperature-Time-Transformation phase diagram are in good agreement with continuous-variable simulations and experiment. 

(2) On quenching, for different material parameters $T_c/T_0, \xi^2, A_1, E_0$, martensite fraction $n_m(t)$ can have slow isothermal and fast athermal conversions. The conversion times $\bar{t}_m$ can transform from rapid athermal to slow isothermal or vice versa on changing the material parameters;  and athermal/isothermal/austenite regime diagrams are obtained in material parameters.

(3) Focusing on the athermal parameter regime, we find rapid conversions below a spinodal like temperature and incubation delay above it, as in the experiment. The conversion-delay times have Vogel-Fulcher divergences, which are insensitive to Hamiltonian energy scales $E_0$ and conversion rates have Log-normal distributions, as in scalar-OP SR transition, from entropy barriers. 

(4) The Temperature-Time-Transformation diagram in the athermal parameter regime has crossover temperatures and are understood through parametrization of domain wall textures by surrogate droplet energies. 

(5) During conversion incubation ($t_m$), evolutions of microstructures, stress distributions and domain-wall thermodynamics are monitored. The initial martensite seeds in the austenite at $t=0$ disappear to form a domain-wall vapour of single variant droplet(s), that incubates before generating $N_v$ variants, one after the other, by autocatalytic twinning to convert to domain-wall liquid. The wandering domain-walls then orient later to a domain-wall crystal. During incubation, stress distributions remain sharply peaked, and there are finite steps in (excess) internal energy, (excess) entropy. 

(6) On switching off the power-law anisotropic potentials, we find no incubations in conversions, no Vogel-Fulcher divergences and the microstructure is multi-slab martensite.

Systematic experiments on athermal martensites can look for martensite formation and growth during conversion incubation and their divergences as well as distributions close to the transition.

Monte Carlo simulations presented in this paper are on 2D strain-pseudospin models for ferroelastic transitions with vector-OP. We also find similar conversion incubation-delays in 3D strain-pseudospin models for tetragonal-to-orthorhombic $(N_v=2, N_{OP}=1)$, cubic-to-tetragonal $(N_v=3, N_{OP}=2)$, cubic-to-orthorhombic $(N_v=6, N_{OP}=2)$, and cubic-to-trigonal $(N_v=4, N_{OP}=3)$ ferroelastic transitions \cite{R21}.

{\it Acknowledgements:} It is a pleasure to thank S.R. Shenoy, T. Lookman and K.P.N. Murthy for very helpful discussions, and S.R. Shenoy for help with the manuscript. The University Grants Commission, India is thanked for Dr. D.S. Kothari Postdoctoral Fellowship.

 \section*{ APPENDIX: internal stresses}

The local internal stresses, $p_{2}(r)=\partial {\bar F} / \partial e_2 (\vec r)$, and $p_{3}(r)=\partial {\bar F} / \partial e_3 (\vec r)$ for TCR transition are obtained as,
$$p_{2}(r)=\epsilon[4\epsilon^{3} S_{2}^3-6\epsilon S_{2}^2+2S_{2}(2\epsilon^2 S_{3}^2+\tau)+ 6\epsilon S_{3}^2$$
$$+ 2 \xi^2 \Delta^2 S_{2}+A_{1}( U_{22} S_{2}+U_{33} S_{3})],~~(A.1a)$$
$$p_{3}(r)=\epsilon[4\epsilon^2 S_{3}^3+2 S_{3}(2\epsilon^2 S_{2}^2+6 \epsilon S_{2}+\tau)+ 2 \xi^2 \Delta^2 S_{3}$$
$$+ A_{1}(U_{23}S_{2}+U_{33}S_{3})].~~(A.1b)$$  

The local internal stresses for SO transition are,
$$p_{2}(r)=\epsilon[2S_{2}(3S_{2}^4-4S_{2}^2+\tau)+2\xi^2 \Delta^2 S_{2}$$
$$+A_{1}(U_{22}S_{2}+U_{33}S_{3})] ,~~(A.2a)$$
$$p_{3}(r)=\epsilon[2S_{2}(3S_{3}^4-4S_{3}^2+\tau)+2\xi^2 \Delta^2 S_{3}$$
$$+A_{1}(U_{23}S_{2}+U_{33}S_{3})].~~(A.2b)$$

The local internal stresses for TO transition are,
$$p_{2}(r)=\epsilon[2S_{2}(3S_{2}^4-4S_{2}^2+\tau)+2\xi^2 \Delta^2 S_{2}$$
$$+A_{1}(U_{22}S_{2}+U_{33}S_{3})],~~(A.3a)$$
$$p_{3}(r)=\epsilon[2S_{2}(3S_{3}^4-4S_{3}^2+\tau)+2\xi^2 \Delta^2 S_{3}$$
$$+A_{1}(U_{23}S_{2}+U_{33}S_{3})].~~(A.3b)$$


\begin{thebibliography}{2004}
\bibitem{R17}
Figure 9 shows the $t_h$-independent data and also $t_h$-dependent conversion times, that are extracted from 'finite size scaling' in $1/t_h \rightarrow 0$ \cite{R14}. The $t_h$-independent linear portions in $1/\log(\bar t_{m})$ versus $\tau$ are extrapolated to intersect the $\tau$-axis, and a temperature $\tau_4$ is defined, that almost matches with the operationally defined $\tau_4$ in Figure 9.
\end{thebibliography}
\end{document}